\documentclass{emulateapj}
\pdfoutput=1
\usepackage{epsfig}
\usepackage{amsmath}
\usepackage{amssymb}
\usepackage{wasysym}
\usepackage{appendix}
\usepackage{verbatim}
\def\beq{\begin{equation}}
\def\eeq{\end{equation}}
\def\ba{\begin{eqnarray}}
\def\ea{\end{eqnarray}}
\def\bal{\begin{align}}
\def\eal{\end{align}}

\begin{document}

\title{Angular Momentum Transport via Internal Gravity Waves in Evolving Stars}

%\author[J. Fuller et al.]{Jim Fuller$^{1,2}$,\thanks{Email:jfuller@caltech.edu}
%Daniel Lecoanet$^{1,3}$,
%Matteo Cantiello$^1$,
%Ben Brown$^1$\\
%\\$^1$Kavli Institute for Theoretical Physics, Kohn Hall, University of California, Santa Barbara, CA 93106, USA
%\\$^2$TAPIR, Mailcode 350-17, California Institute of Technology, Pasadena, CA 91125, USA
%\\$^3$Astronomy Department, B-20 Hearst Field Annex 3411, University of California at Berkeley, Berkeley, CA 94720-3411}

\author{Jim Fuller$^{1,2}$, Daniel Lecoanet$^{1,3}$, Matteo Cantiello$^{1}$, Ben Brown$^{1,4}$}

\affil{$^1$Kavli Institute for Theoretical Physics, University of California, Santa Barbara, CA 93106, USA; jfuller@caltech.edu
\\$^2$TAPIR, California Institute of Technology, Pasadena, CA 91125, USA
\\$^3$Astronomy Department, University of California at Berkeley, Berkeley, CA 94720-3411, USA
\\$^4$Laboratory for Atmospheric and Space Physics \& Department of Astrophysical and Planetary Sciences, University of Colorado, Boulder, CO 80309, USA}

\begin{abstract}

Recent asteroseismic advances have allowed for direct measurements of the internal rotation rates of many sub-giant and red giant stars. Unlike the nearly rigidly rotating Sun, these evolved stars contain radiative cores that spin faster than their overlying convective envelopes, but slower than they would in the absence of internal angular momentum transport. We investigate the role of internal gravity waves in angular momentum transport in evolving low mass stars. In agreement with previous results, we find that convectively excited gravity waves can prevent the development of strong differential rotation in the radiative cores of Sun-like stars. As stars evolve into sub-giants, however, low frequency gravity waves become strongly attenuated and cannot propagate below the hydrogen burning shell, allowing the spin of the core to decouple from the convective envelope. This decoupling occurs at the base of the sub-giant branch when stars have surface temperatures of $T\approx 5500$K. However, gravity waves can still spin down the upper radiative region, implying that the observed differential rotation is likely confined to the deep core near the hydrogen burning shell. The torque on the upper radiative region may also prevent the core from accreting high-angular momentum material and slow the rate of core spin-up. The observed spin-down of cores on the red giant branch cannot be totally attributed to gravity waves, but the waves may enhance shear within the radiative region and thus increase the efficacy of viscous/magnetic torques. 

\end{abstract}

\section{Introduction}

Astronomers have known for hundreds of years that stars rotate. The understanding of {\it how} stars rotate is much less certain. For stars other than the Sun, there have been, until recently, essentially no direct measurements of internal rotation rates. Nor has there developed a comprehensive theoretical understanding of how internal rotation rates change as stars evolve and their structures contort in the continual battle to maintain hydrostatic equilibrium. Such an understanding is essential if we wish to assess the impact of rotation on stellar birth, stellar life, and stellar death.

Recent advances in observational data (most notably due to the superb photometry obtained by the {\it Kepler} satellite) have allowed for asteroseismic measurements of internal stellar rotation rates. By measuring the rotational splitting of mixed modes (stellar oscillation modes with gravity mode character in the stellar core and pressure mode character in the convective envelope) Beck et al. (2012,2014) measured the internal rotation rates of four stars ascending the red giant branch (RGB). Mosser et al. 2012 used similar methods to measure the core rotation rates of many RGB and helium-burning clump stars. Deheuvels et al. (2012, 2014) has used asteroseismic techniques to measure the core and envelope rotation rates of seven sub-giant stars. 

These studies revealed the existence of large amounts of differential rotation in post-main sequence stars, indicating that the inner cores of these stars rotate significantly faster than the envelopes. Throughout this paper, the inner core refers to the g-mode cavity of the sub-giants/red giants, which is mostly localized at and below the hydrogen burning shell overlying the degenerate helium core. The envelope refers to the thick convection zone comprising the bulk of the radial extent of the star. Typical inner core rotation rates for these stars are on the order of ten days, while the envelopes rotate at much longer periods ($P \gtrsim 50$ days). 

Recently, Kurtz et al. 2014 asteroseismically measured the rotation profile of a pulsating A-type main sequence star. They found the data were consistent with a (nearly) rigidly rotating envelope. Moreoever, helioseismic measurements of the radiative core of the Sun indicate it is also nearly rigidly rotating. For the purposes of our investigation, the slight differential rotation in these stars ($\sim 7 \%$ for Kurtz's star, and $\sim 30 \%$ latitudinal differential rotation in the convective envelope of the Sun) is negligible compared to the strong differential rotation ($> 100\%$) observed in more evolved stars. 

The existing measurements paint an interesting picture. Stars appear to maintain nearly rigid body rotation on the main sequence, implying efficient angular momentum (AM) transport mechanisms. At some point after the main sequence, stars begin to develop large amounts of differential rotation as the cores contract and spin up. Intriguingly, the measurements of sub-giant/RGB stars indicate that the cores rotate much faster than they would if the stars were rigidly rotating, but much slower than they would in the absence of AM transport. Therefore, AM transport mechanisms in evolved stars must be acting but are not efficient enough to maintain rigid rotation. 

Several mechanisms have been proposed to produce AM transport within stellar interiors. Rather than summarize them all here, we instead refer the reader to Tayar \& Pinsonneault 2013 and Cantiello et al. 2014 for summaries and references. The basic picture appears to be that convective motions enforce nearly rigid rotation throughout stellar convection zones. In radiative zones, AM transport via waves and/or magnetic fields likely dominates. However, it appears that magnetic mechanisms have trouble producing enough AM transport to match observations (Dennisenkov et al. 2010, Cantiello et al. 2014), and wave-driven transport may therefore be important. 

In a series of papers, several authors (Kumar \& Quataert 1997, Zahn et al. 1997, Kumar et al. 1999, Talon \& Kumar 2002, Talon \& Charbonnel 2005, Charbonnel \& Talon 2005, Talon \& Charbonnel 2008) investigated wave driven AM transport in low mass stars. Internal gravity waves (IGW) are generated by convective motions near the radiative-convective interface. The IGW propagate into radiative regions and deposit their AM where they damp. Most of the IGW damp via radiative diffusion before they are able to reflect and set up stellar oscillation modes. The authors above found that IGW are capable of redistributing AM within Sun-like stars on short timescales and that IGW can partially account for the nearly rigid rotation of the Sun's radiative zone, although other mechanisms may also be required (Denissenkov et al. 2008).

There have also been recent advances in numerical simulations of IGW in stellar interiors. Barker \& Ogilvie 2010 simulated tidally excited IGW propagating and non-linearly breaking near the center of a solar-type star. More recently, Rogers et al. 2013 simulated convectively excited IGW in massive main sequence stars, while Alvan et al. 2014 simulated convectively excited IGW in the Sun. The simulations are impressive, as they globally model convective motions, IGW excitation, and subsequent IGW propagation and dissipation. Their results serve as a basis for comparison with observations and our analytical results.

In this paper, we examine the role of AM transport via convectively excited IGW as stars evolve off the main sequence and up the RGB. In particular, we attempt to determine whether IGW can account for the necessary AM transport in sub-giant/RGB stars to match asteroseismic observations. We find that as stars evolve, the increasing Brunt-V\"{a}is\"{a}l\"{a} frequency in the radiative zone makes it opaque to IGW, preventing the IGW from penetrating into the inner core. On the lower sub-giant branch, the inner core decouples from the influence of the convectively excited IGW, allowing the core to spin-up. The IGW therefore have difficulty in spinning down the cores of RGB stars on their own, although they are still capable of removing large amounts of AM from the outer core. A complete picture of AM transport in these stars may therefore need to account for both IGW and other AM transport mechanisms, e.g., magnetic torques.

Our paper is organized as follows. In Section \ref{basic} we review the basic concepts involved in wave excitation, propagation, dissipation, and AM transport. Section \ref{simple} presents a simple example of how IGW can redistribute AM within the differentially rotating core of an evolved star. In Section \ref{Omstar} we provide some quantitative estimates of IGW characteristics and AM redistribution timescales in different types of stars. We conclude in Section \ref{discussion} with a discussion of our results and their relation to existing observations.

\section{Basic Ideas}
\label{basic}

Here we review some of the basics of IGW generation, propagation, and dissipation. These concepts are also found in previous works (e.g., Kumar \& Quataert 1997, Zahn et al. 1997, Kumar et al. 1999, Talon \& Kumar 2002, Talon \& Charbonnel 2005, Charbonnel \& Talon 2005, Talon \& Charbonnel 2008);
	\!\!\footnote{Kumar \& Quataert 1997 and Zahn et al. 1997 contain a sign error in $m$, causing a fundamental error in the dynamics of prograde vs. retrograde waves. Nonetheless much of the rest of their analysis is quite useful. The sign error was corrected in subsequent works.
	}
here we present only the fundamental aspects crucial to IGW AM transport.

\subsection{Wave Energetics}

Like any other type of wave, IGW transport energy and AM. The waves extract energy/AM from the region of excitation and deposit it in the region of dissipation. In the case of convectively excited waves propagating in the radiative cores of evolving low mass stars, the waves extract AM from the convective zone and deposit it in the radiative interior. 

Convective motions generate waves with an energy flux of order
\beq
\label{Edot}
\dot{E} \sim \mathcal{M} L
\eeq
(Goldreich \& Kumar 1990, Kumar et al. 1999), where $\mathcal{M}$ is the convective Mach number near the radiative-convective interface,
	\!\!\footnote{Lecoanet \& Quataert define the convective Mach number as $\omega_c/N_0$, where $N_0$ is a characteristic Brunt-V\"{a}is\"{a}l\"{a} frequency below the convection zone. We choose to evaluate the convective Mach number as $v_c/c_s$, where $v_c$ is the convective velocity and $c_s$ is the sound speed at the base of the convection zone. The two expressions are the same within a factor of a few, and the uncertainty in the Mach number is smaller than uncertainties due to unknown physics, e.g., the characteristics of convection and the spectrum of IGW that it generates. 
	}
and $L$ is the stellar luminosity. For overlying convective zones, the waves carry an energy flux $\dot{E}$ downwards. For deep convection zones in low mass stars, $\mathcal{M} \ll 1$ and the waves have a negligible impact on the net energy transport. The characteristic angular frequency of the waves is the angular convective turnover frequency $\omega_c$ near the radiative-convective interface, which we calculate via equation 5.51 of Hansen \& Kawaler 1994:
\beq
\omega_c = \frac{v_c}{\lambda},
\eeq
where $\lambda$ is the mixing length and $v_c$ is the convective velocity, which for efficient convection is
\beq
v_c = \bigg( \frac{\lambda g}{\rho c_P T} F \bigg)^{1/3} \simeq \bigg(\frac{F}{\rho}\bigg)^{1/3},
\eeq
and all quantities have their usual meaning. The characteristic AM flux carried by the waves is 
\beq
\label{Jdot}
\dot{J} \sim \frac{m_c}{\omega_c} \dot{E},
\eeq
where $m_c$ is a characteristic azimuthal number associated with the waves, which is typically of order $l \sim m \sim$ several.
	\!\!\footnote{For convection in a slowly rotating star, mixing length theory predicts the energy bearing eddies to have horizontal extent of $\sim H$, where $H$ is a pressure scale height. The excited waves have peak energies where $l\sim r/H$ (Goldreich \& Kumar 1990) which is usually of order $l \sim$ several at the base of convective envelopes.
	}
For a slowly rotating ($\Omega \ll \omega_c$) star, we may expect prograde (positive $m$) waves and retrograde (negative $m$) waves to be excited to equal amplitudes, such that no net AM flux is carried by the waves. As we shall see below, differential rotation naturally produces a wave filter, selectively allowing prograde or retrograde waves to pass through, generating a non-zero net AM flux. 

The AM flux of equation \ref{Jdot} is quite large. The characteristic timescale for waves to change the spin of the radiative region is 
\begin{align}
\label{tchar}
t_{\rm waves} \sim \frac{I_{\rm rad} \Omega}{\dot{J}},
\end{align}
where $I_{\rm rad}$ is the moment of inertia of the radiative zone and $\Omega$ is the angular rotation frequency. For the Sun, sub-giants, and red giants, $t_{\rm waves} \lesssim 10^5$ years. Actual wave spin-up timescales are typically longer (although still much shorter than stellar evolution timescales) because most waves are unable to propagate far into the radiative region. Nonetheless, it is important to realize that IGW are capable of changing the spin of the radiative regions on timescales much shorter than the stellar evolution time scale.

\subsection{Wave Propagation and Dissipation}

\begin{figure}
\begin{center}
\includegraphics[scale=0.43]{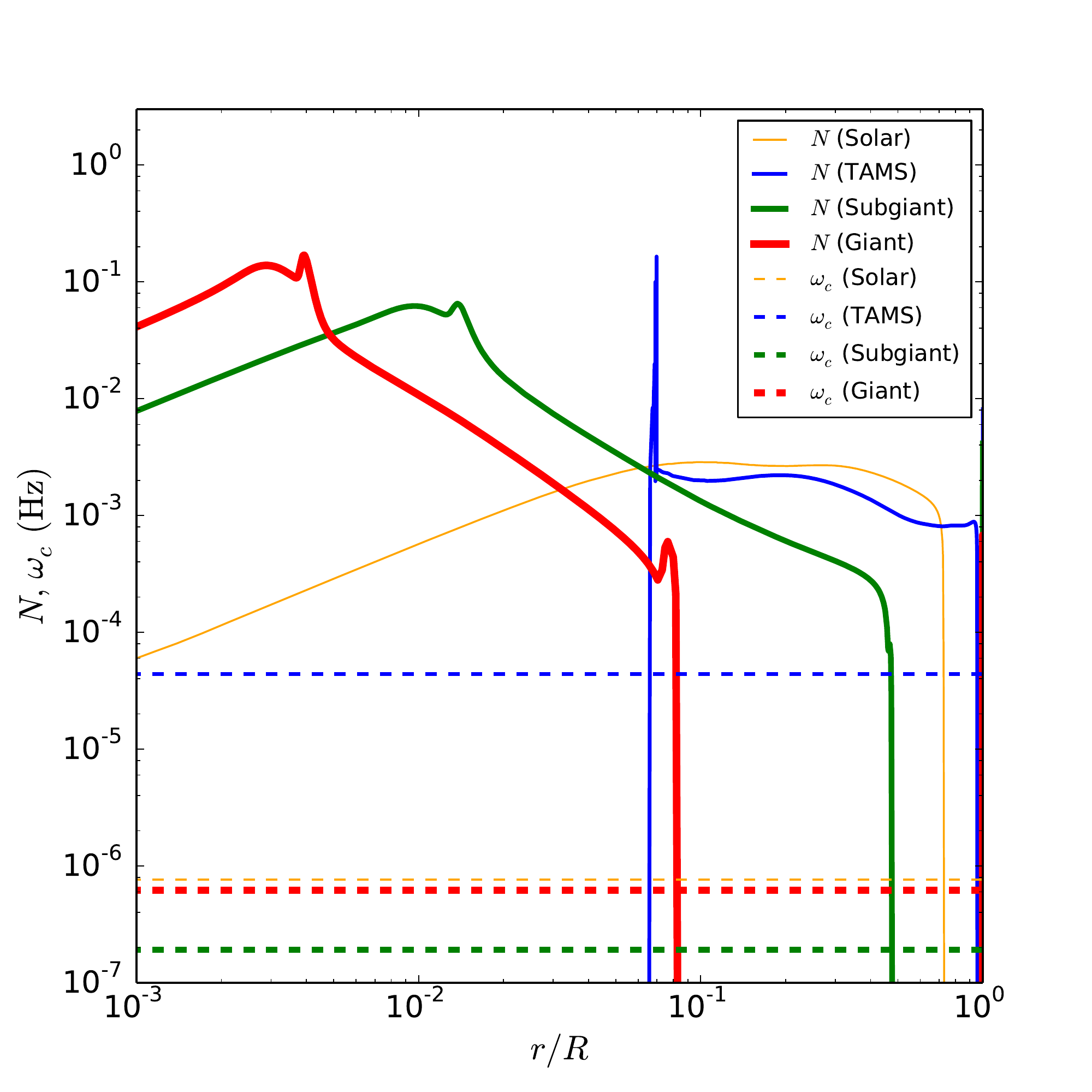}
\end{center} 
\caption{ \label{Prop} Brunt-V\"{a}is\"{a}l\"{a} frequency $N$ as a function of stellar radius for several characteristic stellar models, generated from the MESA stellar evolution code. The horizontal lines are the angular convective turnover frequencies $\omega_c$ near the bottom of the surface convection zones. The masses and radii of the models are $M=M_\odot$, $R=R_\odot$ (Solar), $M=1.5 M_\odot$, $R=1.86 R_\odot$ (TAMS), $M=1.5 M_\odot$, $R= 2.86 R_\odot$ (Subgiant), $M=1.5 M_\odot$, $R=8.0 R_\odot$ (Giant), and all models have solar metallicity.}
\end{figure}

The IGW generated by convection typically have very small frequencies compared to the Brunt-V\"{a}is\"{a}l\"{a} frequencies in the radiative region, i.e., $\omega_c \ll N$ (see Figure \ref{Prop}). Consequently, the IGW have very short wavelengths, and their propagation/dissipation is well approximated by WKB scaling relations.
	%\!\!\footnote{We have not examined the consequences of IGW AM transport on chemical mixing, nor have we included rotational mixing in our stellar models. We expect compositional mixing via IGW to be negligible since the waves damp via radiative diffusion and not turbulent wave breaking. Compositional mixing via rotation in the presence of IGW has been examined in Talon \& Charbonnel 2005, Charbonnel \& Talon 2005. Although rotational mixing may alter surface abundances, we do not expect it to significantly modify the stellar structure.
	%}
	\!\!\footnote{We ignore the effect of magnetic fields on IGW dynamics. In most cases this approximation is justified because wave frequencies $\omega$ are typically larger than Alven frequencies $\omega_A$. However, for strong toroidal fields located in a tachocline, magnetic fields may be important. This possibility has been investigated by Kumar et al. 1999, MacGregor \& Rogers 2010, and Rogers \& MacGregor 2011.
	}
The WKB radial wave number of IGWs is
\beq
k_r^2 = \frac{l(l+1) N^2}{r^2 \omega^2}
\eeq
and the radial group velocity is 
\beq
v_{g,r} = \frac{r \omega^2}{\sqrt{l(l+1)} N}.
\eeq
Lower frequency waves have shorter wavelengths and slower group velocities, making them more prone to damping. The radial wave damping length is (Zahn et al. 1997)
\begin{align}
\label{ldamp}
L_d = \frac{2 r^3 \omega^4}{[l(l+1)]^{3/2} N N_T^2 K },
\end{align}
where $N_T^2=N^2-N_\mu^2$ is the thermal part of the Brunt-V\"{a}is\"{a}l\"{a} frequency ($N_\mu^2$ is the compositional part) and
\beq
\label{thermdiff}
K = \frac{16 \sigma_B T^3}{3 \rho^2 c_p \kappa}
\eeq
is the thermal diffusivity.
	\!\!\footnote{Equation \ref{ldamp} refers to the {\it radial} damping length, i.e., the radial distance the waves propagate before they dissipate. It is different from equation 10 of Rogers et al. 2013, which describes the {\it total} length traversed by a wave before it dissipates. Because gravity waves propagate primarily horizontally, the total damping length is much longer than the radial damping length. The two differ by a factor ($\omega/N$) which is the pitch angle of the spiral traced out by a propagating wave front. Equation \ref{ldamp} (see also equation \ref{tau}) is the appropriate expression for the radial distance waves propagate before they damp out (in linear WKB theory), and equation 17 of Rogers et al. 2013 is incorrect.
	%}$^,$
	%\!\!\footnote{Both Rogers et al. 2013 and Alvan et al. 2014 find that a radial damping length approximately proportional to $\omega^3$ seems to fit better with their simulation results. In the absence of a detailed theoretical explanation of these results, we focus on the predictions of WKB wave propagation.
	}$^,$
	\!\!\footnote{IGW dynamics are modified in double-diffusive convection zones (see Wood et al. 2013 and references therein). These regions are characterized by a negative thermal part of the Brunt-V\"{a}is\"{a}l\"{a} frequency, $N_T^2$, but a positive total Brunt-V\"{a}is\"{a}l\"{a} frequency $N^2$ due to a stabilizing composition gradient. Double-diffusive convection may exist at the base of the radiative envelope in higher mass main sequence stars and at the base of the convection zone in RGB stars. The effect of double-diffusive convection is to enforce $N_T \sim 0$ implying the wave damping length (see equation \ref{ldamp}) becomes very large, i.e., IGW are essentially undamped in double-diffusive convection zones.
	}
The opacity $\kappa$ in the denominator of equation \ref{thermdiff} is the effective opacity due to heat transport by both photons and degenerate electrons. 

Equation \ref{ldamp} demonstrates that low frequency waves have much smaller damping lengths, and are thus unable to propagate far from the convection zone. Additionally, large values of the $N$ create very short damping lengths, preventing waves from propagating into strongly stratified regions. Characteristic damping lengths are $L_d \sim 10^{-2} R_\odot$ for $l=3$, $\omega=\omega_c$ waves just below the solar convection zone. Most waves therefore damp out long before they reach the stellar core. However, waves of frequency $\omega \gtrsim 5 \omega_c$ have damping lengths longer than a solar radius and can therefore penetrate all the way to the center of the Sun.

As stars evolve off the main sequence, their cores contract and the value of $N$ near the hydrogen burning shell increases markedly. The damping lengths of the waves are correspondingly shortened, preventing the waves from propagating to the centers of the stars. IGW are therefore somewhat ineffective at transporting AM into the cores of evolved stars, as we examine in more detail in Section \ref{Omstar}.

\subsection{Wave Filtering}

Differential rotation within a star naturally filters the IGW that propagate within the star. For simplicity, we assume shellular differential rotation at all times.
	\!\!\footnote{Uniform rotation across spherical shells may be maintained by magnetic torques in radiative zones, even if these torques are inefficient at transporting AM in the radial direction, as argued by Spruit 2002. Strong anisotropic turbulence along isobars (Zahn 1992, Maeder \& Meynet 1997) may also give rise to shellular rotation.
	}
The local wave frequency, measured in the co-rotating frame with angular velocity $\Omega(r)$, is
\beq
\label{omega}
\omega(r) = \omega(r_c) - m \big[ \Omega(r)-\Omega(r_c) \big]
\eeq
where $\omega(r_c)$ is the wave frequency when launched from the convective zone and $\Omega(r_c)$ is the rotation frequency of the convective zone. Consider the case in which the interior layers of the star rotate faster than the surface such that $\Omega(r) > \Omega(r_c)$. The prograde waves ($m>0$) are boosted to lower frequencies by the differential rotation, causing their damping lengths to drastically decrease. If $\omega \rightarrow 0$, the waves encounter a critical layer and are completely damped out. In contrast, the retrograde waves are boosted to higher frequencies by the differential rotation. Their damping length drastically increases, allowing them to propagate much further within the star than they would have otherwise.

The differential wave damping produces an imbalance in the net AM flux. The AM flux carried inward by a train of waves launched from an overlying convective zone is 
\beq
\dot{J}(r) = \dot{J}(r_c) e^{-\tau},
\eeq
where the wave optical depth $\tau$ is
\begin{align} 
\label{tau}
\tau &= 2 \int^{r_c}_r \frac{dr}{L_d} \nonumber \\
&= \int^{r_c}_r dr \frac{[l(l+1)]^{3/2} N N_T^2 K }{r^3 \omega^4}.
\end{align}
A factor of 2 change in $\omega$ changes $\dot{J}(r)$ by a factor of $e^{15 \tau}$, a huge factor for strongly damped waves. Small amounts of differential rotation therefore change the wave frequencies enough to generate a huge difference between AM fluxes carried by prograde and retrograde waves.

Differential rotation thus sets up an efficient wave filter: prograde waves are absorbed before they can propagate far into more rapidly rotating layers of a star. Only retrograde waves pass through, meaning the net AM flux into the rapidly rotating layers is negative. When the retrograde waves dissipate, they deposit their negative AM, spinning down the rapidly rotating layers. The star thus evolves toward a state of rigid rotation.

\subsection{Rotational Evolution}

The wave dynamics presented above do not always proceed so simply. One of the main reasons is the ``anti-diffusive" nature of IGW, that can cause IGW to generate shear rather than destroy it. Indeed, there exists no equilibrium rotation rate in the presence of IGW, as waves cause small perturbations in rotation frequency (a small perturbation is defined as $\Delta \Omega \ll \omega/m$) to grow in amplitude. In a rigidly rotating star, a small perturbation in spin rate is amplified on a timescale 
\beq
\label{tgrow}
t_{\rm grow} = \frac{\pi}{3m^2} \frac{\rho r^4 L_d \omega^2}{\dot{E}},
\eeq
for an energy flux $\dot{E}$ carried by waves of frequency $\omega$ and azimuthal number $m$.\footnote{Equation \ref{tgrow} is essentially the same as equation 13 of Kumar \& Quataert 1997, although due to a sign error they mis-interpreted it as a shear damping time scale rather than a growth time scale.} This is essentially the timescale for waves to change the spin rate of a shell of thickness $L_d \ll r$ by an amount $\omega/m$. 

The shear amplification cannot proceed indefinitely. Once the spin rate has changed by $\Delta \Omega = \omega/m$, the differential rotation creates a critical layer that absorbs incoming waves. At this point, the shear can no longer be amplified because waves damp out just before reaching the critical layer. The shear thus moves toward the source of the IGW (Goldreich \& Nicholson 1989). 

IGW AM transport therefore proceeds in two different modes. Small perturbations in spin ($\Delta \Omega \ll {\bar \omega}/{\bar m}$, where ${\bar m}$ and ${\bar \omega}$ are the characteristic pattern number and angular frequency of waves which dominate the AM flux) are amplified on the time scale $t_{\rm grow}$. Large perturbations in spin ($\Delta \Omega \gtrsim {\bar \omega}/{\bar m}$) efficiently filter waves in such a manner as to allow them to reduce the differential rotation until $\Delta \Omega \sim {\bar \omega}/{\bar m}$. Hence, IGW cannot enforce rigid rotation, although they can prevent the build-up of large amounts of differential rotation. The Sun's nearly rigidly rotating radiative zone has $\Delta \Omega \ll \omega_c$ (Howe 2009), which indicates that some other mechanism (e.g., magnetic torques) prevents the build-up of shear (Denissenkov et al. 2008).

\subsection{Complications}

Above, we ignored viscous effects that are important in cases where IGW are able to produce large amounts of shear. Viscosity coupled with IGW-induced shear can produce shear-layer oscillations (SLO) near the base of the convection zone (see Kumar et al. 1999, Kim \& MacGregor 2001, Talon et al. 2002, Talon \& Charbonnel 2005) that may have been detected in the Sun (Howe et al. 2000).\footnote{The phyiscal nature of SLO are essentially the same as the quasi-biennial oscillation observed in the Earth's atmosphere due to upwardly propagating IGW (Shepherd 2000, Baldwin et al. 2001)} The timescale of the SLO is approximately equal to $t_{\rm grow}$ evaluated for $\omega=\omega_c$, and is on the order of years in the Sun. 

For the purposes of the secular evolution of global scale differential rotation, many of the anti-diffusive effects of IGW, such as the SLO, can be ignored. The SLO has a short time scale and likely does not qualitatively affect evolution on longer time-scales. Instead, secular evolution arises from wave filtering due to the steady-state (or averaged) differential rotation. This filtering allows IGW to reduce the differential rotation until its amplitude is of order $\Delta \Omega \sim \bar{\omega}/\bar{m}$. 

An additional complication is that we must include the effects of a broad spectrum of waves (consisting of large ranges in $l$, $m$, and $\omega$), whose shape is not well-constrained (see Section \ref{specsec}) and which has a stochastic nature. The stochastic nature of the wave excitation is likely to average out into a smooth wave spectrum over comparatively long ($t \gg \omega_c^{-1}$) spin-down time scales.
	\!\!\footnote{This may not be true in the diffuse atmosphere of high mass stars where small moments of inertia produce very small wave spin-up time scales and may allow for stochastic evolution of the spin frequency/direction of the atmosphere, see Rogers et al. 2013.
	}
Although the general tendency for waves to reduce large amplitude background differential rotation is not strongly dependent on the wave spectrum, the details of the process can be. 

Finally, we have ignored the influence of the Coriolis force on the IGWs even though the local spin frequency $\Omega(r)$ can be comparable to or greater than the local wave frequency $\omega(r)$. We expect the effect of rotation on propagating IGWs to be well captured by the traditional approximation (Bildsten et al. 1996, Lee \& Saio 1997), which has been explored in previous works (Pantillon et al. 2007, Mathis et al. 2008, Mathis 2009, Mathis et al. 2013). The main effect of the Coriolis force is to increase the effective value of $l$ for the IGW, decreasing their damping length. This will introduce some quantitative corrections to our findings, although uncertainties in the wave spectrum are likely to be more important. It is also possible that rotation will change the nature of convection and the spectrum of IGW that it generates (Mathis et al. 2014), however since even the non-rotating spectrum is poorly understood, we ignore this issue in this work.

Our goal is to obtain general results that are robust against the details of the effects above. We proceed with a simplified analysis that produces order of magnitude estimates for secular wave spin-down time scales, and defer a more precise description of AM redistribution via IGW to future work.

%Additionally, waves cause fluctuations in rotation to be transported toward the source of the waves (this produces the outwardly propagating differential  rotation fronts seen in Talon et al. 2002, Talon \& Charbonnel 2005, Charbonnel \& Talon 2005, Mathis et al. 2013), meaning that the net effect of convectively excited IGW is to advect differential rotation toward the convective zone, where it will be absorbed into the convective envelope.

\section{Simple Example}
\label{simple}

\begin{figure}
\begin{center}
\includegraphics[scale=0.45]{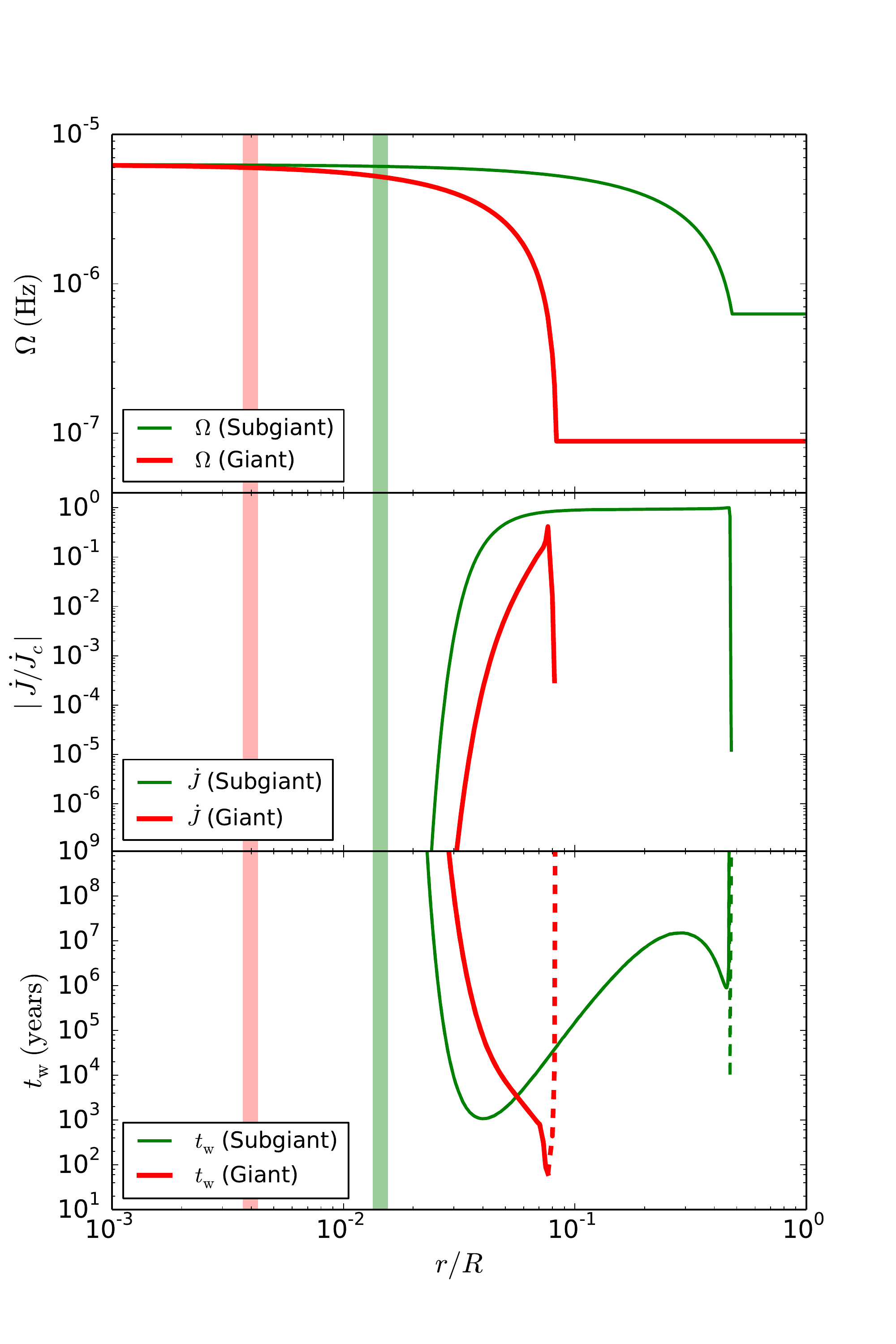}
\end{center} 
\caption{ \label{Omlin} Angular momentum transport in differentially rotating stars for the simplified wave spectrum of equation \ref{Jlin}. {\bf Top:} Linear differential rotation profile in a sub-giant and red giant model (with same parameters as in Figure \ref{Prop}) calculated via equation \ref{omegalin}. The corresponding rotation periods are $P(r=0) = 10$ days, $P(r=R) = 100$ days (sub-giant), and $P(r=R) \approx 800$ days (giant). {\bf Middle:} Absolute value of total AM flux through the surface at radius $r$, in units of the flux launched from the convective zone $\dot{J}_c$ (see text).  {\bf Bottom:} Wave spin-down timescale $t_{\rm w}$ as a function of $r$. The dashed portions of the curves indicate regions which are being spun-up by dissipating prograde waves, while the solid portions are being spun down by dissipating retrograde waves. The shaded vertical columns indicate the location of the hydrogen burning shell in the like-colored model. In this simple example the differential rotation of the outer radiative region is reduced on short time scales, while the inner core is nearly unaffected. }
\end{figure}

Asteroseismic analyses (Beck et al. 2012,2014; Deheuvels et al. 2012,2014; Mosser et al. 2012) of sub-giant and RGB stars reveal that the stellar cores rotate much faster than the envelope. Unfortunately the measurements are not able to provide precise angular velocity profiles. Let us consider the simple case in which the angular velocity $\Omega(r)$ varies linearly with radius between the core and the envelope, such that 
\beq
\label{omegalin}
\Omega(r) = \Omega(r_c) + \frac{r_c - r}{r_c} \big[ \Omega(r_g)-\Omega(r_c) \big],
\eeq
where $\Omega(r_g)$ is the rotation rate of the g-mode cavity from asteroseismic measurements. Example rotation profiles for a sub-giant and red giant are shown in Figure \ref{Omlin}. 

This angular velocity profile will evolve in the presence of convectively excited IGW. Consider a simplified wave spectrum consisting only of $l=3$, $m=\pm3$ waves with $\omega= 2 \omega_c$, and with total energy flux $\dot{E}_c = 10^{-2} \mathcal{M} L$. We have reduced the energy flux from equation \ref{Edot} to account for the lower energy contained in higher frequency waves (see Section \ref{specsec}). Since we have not included a full IGW spectrum, this  exercise provides an approximate lower limit for IGW fluxes (and an upper limit for IGW spin-down times). The total AM flux into any region of the star is given by
\beq
\label{Jlin}
\dot{J} = \frac{3}{2\omega_+} \dot{E}_c e^{-\tau_+} - \frac{3}{2\omega_{-}} \dot{E}_c e^{-\tau_{-}}.
\eeq
Here, the $+$ and $-$ subscripts refer to the prograde and retrograde waves, respectively. The local wave frequencies $\omega$ are calculated from equation \ref{omega}, while the optical depths $\tau$ are calculated from equation \ref{tau}.

In our scenario, the prograde waves encounter a critical layer (where $\omega \rightarrow 0$) only a small distance below the convective zone, and are completely absorbed. This absorption may contribute to shear-layer oscillations localized on short time scales (see above) that are localized near the radiative-convective interface, but we ignore this issue here. The retrograde waves are boosted to higher frequencies by the differential rotation and carry their AM deep into the radiative zone. They dissipate above the hydrogen burning shell where the rise in $N$ lowers their damping length. 

The net AM flux $\dot{J}$ is shown in the middle panel of Figure \ref{Omlin}. The value of $\dot{J}/\dot{J}_c$ is zero at the base of the convective zone because there is an equal flux of prograde and retrograde waves. It quickly rises to $|\dot{J}/\dot{J}_c| \simeq 1$ because the prograde waves are absorbed at the critical layer slightly below the convective interface. The value of $\dot{J}/\dot{J}_c$ falls off deeper in the star as the retrograde waves damp out.

We also plot the wave spin-down timescale
\beq
t_{\rm w}(r) = \frac{-\Omega_s dI/dr}{d\dot{J}/dr},
\eeq
where $I$ is the total moment of inertia of layers interior to radius $r$, and $dI/dr= (8\pi/3)\rho r^4$. It is evident from Figure \ref{Omlin} that the waves will change the spin of the star on very short timescales, with $t_{\rm w}(r)$ as short as $\sim 50$ years in the case of the red giant. We can thus conclude that the linear differential profile adopted for this example is a very unstable configuration and would be wiped out on time scales much shorter than the stellar evolution time scales. Most of the differential rotation in real stars must be confined to the inner part of the core ($r/R \lesssim 3\times10^{-2}$) where $t_{\rm w}(r)$ is longer than the stellar evolution time scales. This conclusion is consistent with the results of Deheuvels et al. 2014, whose asteroseismic inversions show a tentative preference for differential rotation restricted to layers near the hydrogen burning shell.

\section{Angular Momentum Transport in Evolving Stars}
\label{Omstar}

The simple example above highlighted that IGWs will likely confine rapid rotation to well below the radiative-convective interface in evolved stars. In this section we generalize our results for more realistic frequency spectra. However, the goal is still to obtain simple results which are not strongly dependent on the details of the wave spectrum.

In the analysis below, we will consider waves propagating through a rigidly rotating radiative zone. If low frequency IGW are able to propagate into regions of significant differential rotation they can wipe it out on very short timescales, as shown above. Therefore regions permitting large wave energy fluxes should not contain strong differential rotation. Our main goal is then to determine which regions of the star are transparent to waves in the absence of differential rotation.

\subsection{Spectrum of Convectively Generated Internal Gravity Waves}
\label{specsec}

There is broad agreement that convective motions most efficiently generate IGW when the length scales and timescales of the convection and the IGW are comparable (Lighthill 1978). The dominant source of IGW are large-scale convective rolls with size $H$ and with coherence times $\omega_c^{-1}$.  This generates waves with frequencies $\omega \sim \omega_c$, horizontal mode number $l \sim r_c/H$, and radial wavenumber $(N/\omega_c) H^{-1}\gg H^{-1}$, where $N$ is a typical Brunt-V\"{a}is\"{a}l\"{a} frequency in the radiative zone.  The luminosity of these low frequency waves is 
\beq
\label{edoteta}
\dot{E} = \eta \mathcal{M} L.
\eeq
Here, $\eta$ is an efficiency factor of order unity. This is the prediction of e.g., Press 1981, Garcia Lopez \& Spruit 1991, Goldreich \& Kumar 1990, Kumar, Talon \& Zahn 1999, Lecoanet \& Quataert 2013, and is consistent with recent numerical simulations (Alvan et al. 2014).

Although the peak of the excitation spectrum is well understood, the rest of the spectrum is poorly constrained.  The primary difficulty is that high frequency waves are excited by small length scale convective motions, which are difficult to resolve in simulations or experiments.  For instance, the power spectra of convective motions presented in the simulations of Belkacem et al. 2009 and Alvan et al. 2014 using the ASH code (Clune et al. 1999, Brun et al. 2004), look very different from the power spectra measured in far more turbulent experiments (e.g., Niemela 2000).  Even theoretically, there is no consensus on whether the small scale motions follow a Kolmogorov (Kolmogorov 1941) energy cascade, or a Bolgiano-Obukhov (Bolgiano 1959, Obukhov 1959) entropy cascade (e.g., Lohse \& Xia 2010).  Note, however, that the most turbulent simulations and experiments suggest that small scale fluctuations follow a Kolmogorov cascade (e.g., Boffetta 2012, Lohse \& Xia 2010 and references within).

Because direct numerical simulations of wave excitation by convection (Rogers \& Glatzmaier 2005, Brun et al. 2011, Rogers et al. 2013, Alvan et al. 2014) may not have fully resolved the turbulent motions generating high frequency waves, we instead turn to theoretical predictions based on the assumption of a Kolmogorov power spectrum of convective motions.  Lecoanet \& Quataert 2013 predict a wave power spectrum
\beq
\frac{d \dot{E}(\omega,l)}{d\omega \ d l} \sim \frac{\dot{E}}{l \omega_c} \left(\frac{\omega}{\omega_c}\right)^{-a} \left(l \frac{H}{r_c}\right)^b \left(l \frac{d}{r_c}\right)^c
\eeq
where $d$ is the width of the transition regime between the radiative and convective regions, and $\dot{E}$ is the total wave energy flux given by equation \ref{edoteta}. Here, $k_{\perp} = \sqrt{l(l+1)}/r$ is assumed to be less than $H^{-1}(\omega/\omega_c)^{3/2}$, and $a, b, c$ are power-law coefficients.  Depending on the details of the transition region, Lecoanet \& Quataert 2013 find $a$ between 7.5 and 8.5, $b=4$, and $c$ between 0 and 1. Goldreich \& Kumar 1990 predict $a=7.5$ and $b=3$.

There is a sharp decline in wave luminosity with frequency because only small eddies have high frequencies, and there is very little power in the small eddies.  Furthermore, the waves most efficiently excited by these small eddies have small horizontal wave lengths, and thus damp very quickly.  The least damped waves have small $l$, and are excited due to the (low probability) coherent superposition of many small eddies (Garcia Lopez \& Spruit 1991).  These waves have
\beq
\label{spectrum}
\frac{d \dot{E}(\omega)}{d\omega} \sim \frac{\dot{E}}{\omega_c} \left(\frac{\omega}{\omega_c}\right)^{-a}.
\eeq
We allow $a$ to be a free parameter, and expect most probable values to lie in the range $3.5 \lesssim a \lesssim 7.5$.

Up to this point, we have not considered the effects of stratification.  Kumar, Talon \& Zahn (1999) suggest that the stratification of a convection zone above a radiative zone can enhance excitation of high frequency waves.  Because the scale height decreases with increasing radius, they argue that the energy-bearing convective motions will shift to smaller length scales and higher frequencies with increasing radius.  Under these assumptions, high frequency, low $l$ waves can be excited by the coherent superposition of many small, energy-bearing eddies.  This allows for much more efficient excitation of high frequency waves.  Their analytic calculations predict a wave spectrum with $a\sim 3.3$, whereas semi-analytic work has suggested $a\sim 4.5$ (see also Talon et al. 2002, Denissenkov et al. 2008). Stratification will not enhance the excitation of high frequency waves for convection zones below radiative zones.

However, recent high resolution simulations of strongly stratified convection do not show this shift of the energy-bearing motions to smaller scales -- rather, they find that the kinetic energy is peaked at large scales throughout the convection zone (Hotta et al. 2014).  If convection in stars is dominated by motions much larger than the local scale height, then it is unlikely that stratification will amplify the excitation of high frequency waves. In this work, we adopt $a=4.5$ as a fiducial value, but we caution that both steeper (larger $a$) and shallower (smaller $a$) frequency spectra are certainly possible.

Requiring $\int^\infty_{\omega_c} \dot{E}_\omega d \omega = \eta \mathcal{M} L$ yields the wave energy flux per unit frequency
\beq
\dot{E}_\omega \sim  \frac{a-1}{\omega_c} \bigg(\frac{\omega}{\omega_c}\bigg)^{-a} \eta \mathcal{M} L.
\eeq
The total AM flux of waves with azimuthal number $m$ and frequency near $\omega$ is 
\beq
\label{Jomega}
\dot{J}(r_c,\omega) \sim  \frac{m}{\omega_c} \bigg(\frac{\omega}{\omega_c}\bigg)^{-a} \eta \mathcal{M} L.
\eeq 
Equations \ref{spectrum}-\ref{Jomega} only apply for IGW with $\omega \gtrsim \omega_c$, and they are valid at the radiative-convective interface $(r=r_c)$. Further into the radiative zone where $r<r_c$, the wave spectrum will shift toward higher frequencies since lower frequency waves damp on short length scales.

\subsection{Wave Transport}

The strong dependence of wave optical depth on frequency (equation \ref{tau}) implies the frequency of waves which dominate the energy/AM flux at a given radius is sharply peaked at a characteristic value, $\omega_*(r)$. At the radiative-convective interface, $\omega_*(r_c) \simeq \omega_c$. Below the interface, $\omega_*$ is set by waves whose optical depth is of order unity at that location. Lower frequency waves have been attenuated and higher frequency waves carry less AM via equation \ref{Jomega}. To solve for $\omega_*(r)$, we find the peak in the value of
\beq
\label{Jstar1}
\dot{J}(r,\omega) = \dot{J}(r_c,\omega) e^{-\tau(r,\omega)}.
\eeq
Taking the derivative of equation \ref{Jstar1} with respect to $\omega$ and setting equal to zero yields
\beq
\tau_{*} = \frac{a}{4}.
\eeq
Then, using equation \ref{tau} we find
\beq
\label{omstar1}
\omega(r) = \bigg[\frac{4}{a} \int^{r_c}_r dr \frac{[l(l+1)^{3/2} N_T^2 N K }{r^3} \bigg]^{1/4}.
\eeq
The right hand side of equation \ref{omstar1} depends primarily on the stellar structure with only a very weak dependence on the wave spectrum. Equation \ref{omstar1} applies for frequencies above the convective turnover frequency $\omega_c$, and an expression valid at all frequencies is 
\beq
\label{omstar2}
\omega_*(r) = {\rm max} \bigg[ \omega_c \ , \ \bigg(\frac{4}{a} \int^{r_c}_r dr \frac{[l(l+1)^{3/2} N_T^2 N K }{r^3} \bigg)^{1/4} \bigg] .
\eeq

The AM flux carried by waves with frequency $\omega \approx \omega_*$ is, using equation \ref{Jomega},
\beq
\label{Jstar}
\dot{J}_*(r) \sim \frac{ m \eta \mathcal{M} L}{\omega_c} \bigg(\frac{\omega_*(r)}{\omega_c}\bigg)^{-a} e^{-a/4}.
\eeq
Because the AM flux is dominated by waves with frequency $\omega_*(r)$, equation \ref{Jstar} gives an approximate value for the total AM flux carried by IGW at radius $r$. Since low $l$ waves have the longest damping lengths, equation \ref{Jstar} should be evaluated using $l\sim |m| \sim 1$. The associated spin-up timescale for regions below a radius $r$ is
\beq
\label{tstar}
T_{*}(r) = \frac{\Omega_s I(r)}{\dot{J}_*(r)}.
\eeq

$T_*(r)$ indicates the time scale on which waves could change the spin rate of the region below radius $r$, if the surface of this region contains differential rotation that creates a wave filter. It is different than the spin evolution time scale $t_{\rm w}(r)$ that indicates the wave spin-down time scale of a spherical shell at radius $r$. Unfortunately, $t_{\rm w}(r)$ is strongly dependent on the stellar rotation profile, which is generally unknown. We find that $T_*(r)$ is a better diagnostic because it provides an estimate for the time scale on which rotation rates at radii below $r$ {\it could} change, in the presence of a wave filter at radius $r$. For our purposes, the most important quantities are the maximum value of $T_*(r)$ and its value at the top of the g-mode cavity (for sub-giants and red giants these values are approximately the same). Since the AM flux of equation \ref{Jstar} is somewhat dependent on the slope of the frequency spectrum, we cannot expect equation \ref{tstar} to yield exact results. Nonetheless, it provides a simple method of estimating IGW spin evolution timescales.

We would like to compare the wave synchronization timescale to the spin-up timescale due to stellar evolution. This timescale is found from the spin evolution in the absence of AM transport:
\beq
\frac{d}{dt} \big( I \Omega_s \big) = \Omega_s \dot{I} + \frac{I \Omega_s}{t_s} = 0,
\eeq
which entails a spin up timescale of 
\beq
\label{ts}
\tau_{\rm spin}(r) = - \frac{r}{2 \dot{r}},
\eeq
where $\dot{r}$ is the rate of change of the radius of the mass contained in a spherical shell at $r$. 

\begin{figure*}
\begin{center}
\includegraphics[scale=0.55]{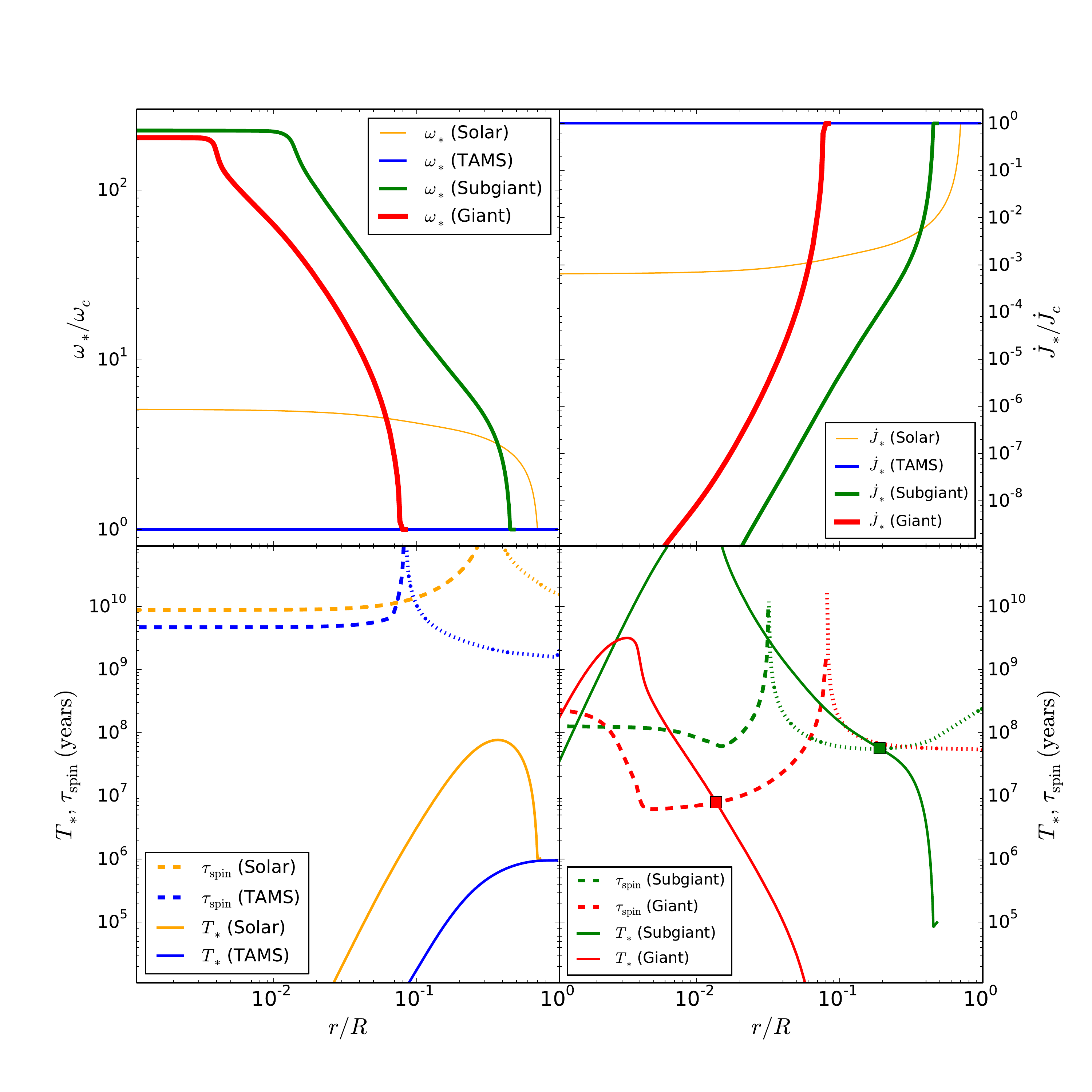}
\end{center} 
\caption{ \label{Omstarfig} {\bf Top Left:} Value of $\omega_*$ from equation \ref{omstar2} as a function of radius, in units of the convective turnover frequency at the base of the surface convection zone, $\omega_c$. This figure was generated using the wave spectrum from equation \ref{Jomega} with $a=4.5$ and $\eta = 0.1$, and using waves of $l=|m|=1$. The stellar models are the same as in Figure \ref{Prop}. {\bf Top Right:} Associated angular momentum flux $\dot{J}_*$ from equation \ref{Jstar}, in units of the flux launched from the convective zone $\dot{J}_c$. {\bf Bottom Panels:} Spin-up time $\tau_{\rm spin}$ calculated from equation \ref{ts}, and wave spin-down time $T_*$ calculated from equation \ref{tstar}. Dotted portions of $\tau_{\rm spin}$ indicate layers which are expanding and spinning down, while dashed portions are contracting and spinning up. In the left panel, the small spin-down times ($T_* \ll |\tau_{\rm spin}|$ in the solar and TAMS models) suggest the radiative interiors are well-coupled to the convective envelopes. In the right panel, the large spin-down times of the sub-giant/red giant models imply the core and envelope are decoupled. The squares mark the radial location of decoupling, i.e., the value of $R_{dc}/R$ (see Figure \ref{EvolveM}). } 
\end{figure*}

Figure \ref{Omstarfig} shows the values of $\omega_*(r)$, $J_{*}(r)$, $T_*(r)$, and $\tau_{\rm spin}(r)$ for a few different stellar models, using $a=4.5$. We have used $l=m=1$ and $\eta=0.1$ as an estimate for the wave energy carried by these limited values of $l$ and $m$. For solar-like stars, the value of $\omega_*(r)$ reaches a maximum of $\sim 5 \omega_c$. Although this reduces the value of $J_*(r)$ to $\sim 10^{-3}$ its value at the convection zone, the remaining AM is still capable of changing the spin frequency on timescales of $\sim 10^8$ years. Even for steep frequency spectra $a \sim 7.5$, the wave timescales are less than the age of the Sun. We therefore agree with previous results (Talon \& Charbonnel 2005, Charbonnel \& Talon 2005) which have found that IGW affect the solar angular velocity profile on short timescales.

Next, we examine the results for a $1.5 M_\odot$ terminal age main sequence (TAMS) star. A star of this mass develops a surface convection zone as it begins to evolve off the main sequence towards cooler surface temperatures. When this convection zone first forms, it is relatively shallow, although it still extends several scale heights and carries nearly all the stellar flux.
	\!\footnote{ The shallow convection zone may inhibit the excitation of large scale (small $l$) waves due to the limited size of convective eddies. In this particular example, our results are not significantly changed even if we use $l=20$ instead of $l=1$ in our calculations. However, in some cases involving shallow convection zones, this issue may be pertinent.}
Because the bottom of the convection zone exists at low densities where the scale height is small, the convective turnover frequency at its base is quite large. The waves generated by the convection therefore have high frequencies and are easily capable of traversing the entire radiative zone (in this case they reflect at the core convection zone and form standing oscillation modes). Moreover, because the convective Mach numbers are larger near the surface convection zone than in the convective core, we expect the surface-generated IGW to dominate the angular momentum flux.
	\!\footnote{ During the main sequence (before the surface convection zone has formed), the core-generated IGW may be important. For our 1.5 $M_\odot$ TAMS model, our calculations indicate that most core-driven waves are damped before making it far into the radiative zone. Wave spin-up time scales are generally longer than the main sequence life time in the bulk of the radiative zone, but become much shorter near the surface due to the small associated moment of inertia (see Rogers \& Lin 2012, Rogers et al. 2013). In some cases the core-driven waves/modes may be observable (Shiode et al. 2013), although here we do not investigate core-driven waves in detail.}
These waves are capable of redistributing AM on short timescales ($T_* \sim 10^6 {\rm yr}$), allowing more massive stars $M \gtrsim 1.4 M_\odot$ to undergo rapid spin evolution as they initially evolve off the main sequence. This situation was also noted in Talon\ \& Charbonnel 2008.

The results are much different for sub-giants and red giants. In these stars, the large values of $N$ near the hydrogen burning shell result in large values of $\omega_*$. Only high frequency waves (relative to the convective turnover frequency) are capable of penetrating into the g-mode cavities probed by asteroseismic measurements. These waves only carry small amounts of AM, assuming the frequency spectrum is reasonably steep ($a \gtrsim 3$). Consequently, the IGW which are able to propagate into the core cannot change its spin on short timescales, and $T_* \gg \tau_{\rm spin}$ near the cores of these stars. Therefore, IGW on their own are likely not capable of efficiently spinning down the cores of ascending RGB stars. This result is reassuring, as the observed rapid core rotation in sub-giants and red giants indicates IGW have not been able to spin down the cores of these stars. However, we note that in the upper radiative zone ($r/R \gtrsim  10^{-1}$ for the sub-giant model and $r/R \gtrsim 2 \times 10^{-2}$ for the red giant model) the wave spin-down timescales are short $(T_* \ll \tau_{\rm spin})$, implying that IGW can still affect the spin of these regions of the star.

\subsection{Wave Decoupling}

The results presented above indicate that IGW can help reduce differential rotation for stars leaving the main sequence, but they cannot keep the inner core (regions at and below the hydrogen burning shell) synchronized as the star evolves up the sub-giant/red giant branch. We would like to know the moment in the evolution at which the waves can no longer penetrate into the core, allowing it to decouple from the surface convection zone.

To determine the epoch of decoupling, we find the stellar evolutionary state at which IGW spin evolution time scales become longer than stellar evolution time scales. We generate stellar models with MESA (Paxton et al. 2011,2013) and evolve them from the zero age main sequence, calculating profiles of $T_*(r)$ and $t_{s}(r)$ at each step. We then find the first stellar model that contains a location below the surface convection zone where $T_*(r) > \tau_{\rm spin}(r)$, and we define this to be the moment of decoupling.

\begin{figure}
\begin{center}
\includegraphics[scale=0.45]{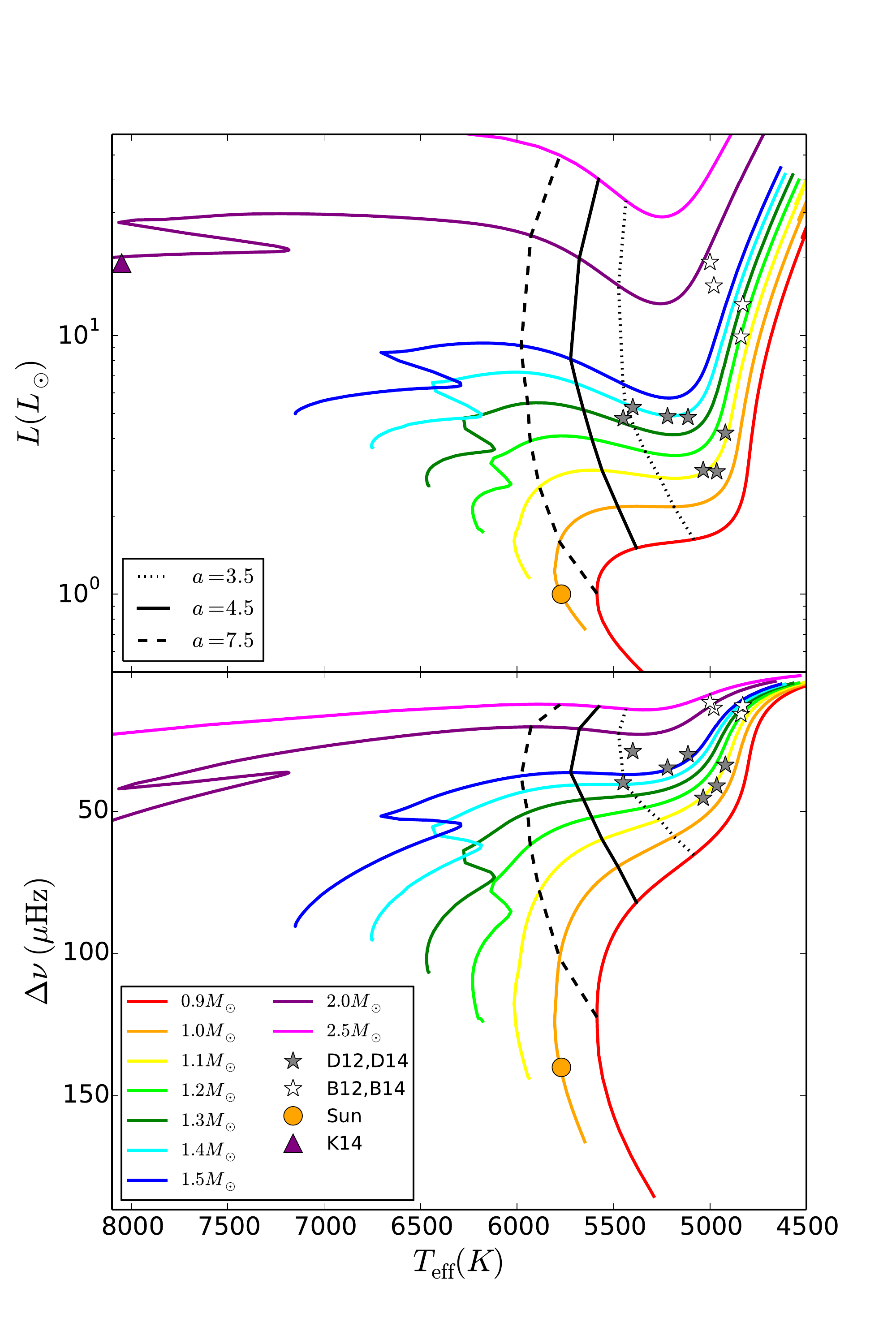}
\end{center} 
\caption{ \label{HRdecoup} {\bf Top:} HR diagram showing the evolutionary tracks for stars of different masses, with solar metallicity. The black lines indicate the moment of decoupling for different frequency spectra parameterized by $a$ (see equation \ref{Jomega}). This plot was made using $l=|m|=1$, and $\eta=0.1$. Stars to the left of the black lines are expected to rotate nearly rigidly, while stars to the right of the black lines can develop large amounts of differential rotation. We have also included the location of the Sun, the A-type star KIC 11145123 anlayzed by Kurtz et al. 2014 (K14), the seven sub-giants analyzed by Deheuvels et al. 2012,2014 (D12,D14) and the four RGB stars analyzed by Beck et al. 2012,2014 (B12,B14). {\bf Bottom:} HR diagram, but with luminosity replaced by the large frequency separation $\Delta \nu$. A large frequency separation is not listed in K14.}
\end{figure}

Figure \ref{HRdecoup} shows evolutionary tracks for stars of different mass and indicates the moment of decoupling for each model. We calculate $T_*$ using waves of $l=|m|=1$, $\alpha=4.5$, and $\eta=0.1$. For stars in the mass range $M_\odot < M < 1.5 M_\odot$ which comprise most of the observed sub-giant/ascending red giant branch {\it Kepler} sample (Schlaufman \& Winn 2013), the moment of decoupling occurs at effective temperatures $T_{\rm eff} \approx 5500K$. At decoupling, the radii of the stars are typically $\sim 1.75$ their main-sequence radius for $M_\odot \lesssim M \lesssim 1.5 M_\odot$. The large frequency separation of stars, $\Delta \nu$, is approximately half its main sequence value at the time of decoupling (see bottom panel of Figure \ref{HRdecoup}). 

The actual stellar parameters at decoupling will depend on stellar metallicity, spin frequency, wave spectrum, etc., but should typically occur in the $5200 {\rm K} \lesssim T_{\rm eff} \lesssim 6200 {\rm K}$ temperature range. In particular, low metallicity stars (such as KIC 7341231, analyzed in Deheuvels et al. 2012) have larger $T_{\rm eff}$ for the same mass, and will have correspondingly warmer temperatures at decoupling. For steep wave spectra ($a \sim 7.5$) the decoupling occurs earlier in the stellar evolution, very soon after core hydrogen exhaustion, and at larger $T_{\rm eff}$. Shallow wave spectra ($a\lesssim3$) do not decouple until later in the stellar evolution, further up the RGB, at evolutionary stages beyond those of the sub-giants observed by Deheuvels et al. (2012,2014).

The decoupling of the core occurs for three reasons. First, the stellar evolution timescales decrease from $\sim 10^9$ years to $\sim 10^7$ years as a star evolves from the main sequence to the RGB, meaning waves have to act on shorter timescales to keep up with the spin-up of the contracting core. Second, as stars evolve across the sub-giant region, their surface convective zone deepens, penetrating further into the star where the convective turnover frequencies are smaller. The frequencies of the convectively excited waves correspondingly decrease, meaning they cannot propagate as far into the core. Third, as the core contracts, its peak Brunt-V\"{a}is\"{a}l\"{a} frequency $N$ increases by over an order of magnitude. The inner core therefore becomes optically thick to the waves, prohibiting efficient core-envelope coupling. 

\begin{figure}
\begin{center}
\includegraphics[scale=0.45]{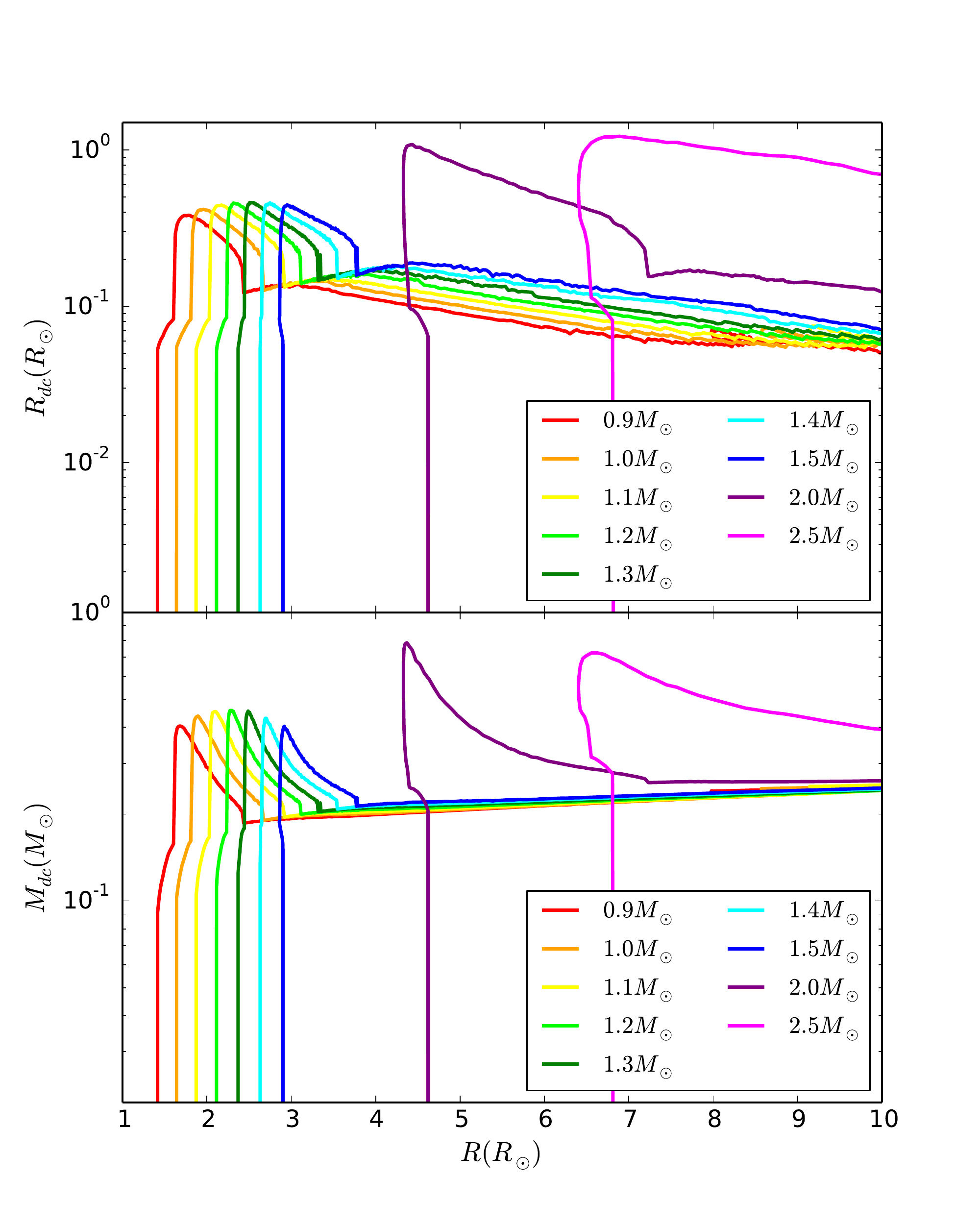}
\end{center} 
\caption{ \label{EvolveM} {\bf Top:} Radius of the decoupled core $R_{dc}$, as a function of the stellar radius $R$ as stars ascend the red giant branch. This figure uses the same parameters as Figure \ref{HRdecoup}, and with $a=4.5$. {\bf Bottom:} Mass of the decoupled core $M_{dc}$. After the onset of decoupling, the mass of the decoupled region is $M_{dc} \approx 0.2 M_\odot$ and is only weakly dependent on the stellar mass and evolutionary state for $M \lesssim 2 M_\odot$.}
\end{figure}

The findings above appear to be consistent with asteroseismic measurements. The sub-giant stars studied by Deheuvels (2014) have temperatures of $T_{\rm eff} \sim 5000 {\rm K}$ and radii in the range $2-3 R_\odot$ range, and therefore have likely evolved past the moment of decoupling. Indeed, Deheuvels (2014) find that these stars appear to have cores which are spinning up with time. Our findings are also consistent with those of Tayar \& Pinsonneault (2013) who find that the internal rotation rate of a low mass sub-giant (KIC 7341231, Deheuvels et al. 2012) requires decoupling to occur at stellar radii of $R\sim1.5-1.9 R_\odot$.

Finally, we can estimate the extent of the decoupled core by searching for the first radial location at which $T_*(r) > \tau_{\rm spin}(r)$ as one travels from the convective envelope inwards (marked by a square in Figure \ref{Omstarfig}). Regions below this radial location are decoupled from convectively excited IGW. Figure \ref{EvolveM} shows the extent of the decoupled region in terms of radius $R_{dc}$ and mass $M_{dc}$ as a function of the stellar radius as stars evolve up the RGB. Typically only the inner part of the radiative core is decoupled, with $M_{dc} \approx 0.2 M_{\odot}$ as the star evolves up the lower RGB. We therefore expect any differential rotation to be restricted to mass coordinates $M(r) < M_{dc} \approx 0.2 M_\odot$.  The decoupled region includes the helium core, the hydrogen burning shell, and a small fraction of the radiative envelope. Steeper wave spectra have larger decoupled regions (comprised by the bulk of the radiative region) while shallower wave spectra have smaller decoupled regions (but still including the helium core and hydrogen burning shell).

\section{Discussion and Conclusions}
\label{discussion}

We have examined the impact of internal gravity waves (IGW) on angular momentum (AM) transport and the internal rotation rates of evolving low mass ($0.9 M \odot \lesssim M \lesssim 2.5 M_\odot$) stars. Convection zones generally excite IGW that carry large fluxes of AM, and the presence of large scale differential rotation within the star sets up an efficient wave filter. The ensuing propagation and dissipation of the filtered waves tends to reduce the differential rotation until its magnitude is comparable to local IGW frequencies.  Therefore, as long as IGW are able to propagate from the convection zone to a region of strong differential rotation, they can reduce differential rotation on timescales much shorter than stellar evolution time scales.

In low mass stars with deep convective envelopes, most of the convectively excited IGW are radiatively damped before they can propagate to the center of the star. Therefore only IGW with frequencies somewhat larger than convective turnover frequencies can affect the stellar core.  For the most plausible IGW spectra (see Section \ref{specsec}), we find that IGW can change main sequence internal rotation rates on short time scales $(T_* \lesssim 10^{8} \ {\rm years})$. However, as stars evolve off the main sequence their cores contract and become opaque to IGW. The core decouples from the envelope, allowing large amounts of differential rotation to develop. 

Our theory is consistent with IGW providing the bulk of AM transport within stars. All stars younger than decoupling, which occurs near $T_{\rm eff} \approx 5500 {\rm K}$ (see Figure \ref{HRdecoup}), have been measured to have small amounts of internal differential rotation. All low mass ($M \lesssim 2.5 M_\odot$) sub-giant/red giant stars older than decoupling have been measured to have large amounts of internal differential rotation. IGW may therefore provide the bulk of the synchronizing torque in low mass stars, although other AM transport mechanisms are likely required to enforce the rigid rotation of the radiative region of the Sun (Denissenkov et al. 2008), and to produce the small degree of spin down observed for RGB cores (Mosser et al. 2012).

%The resultant timescale for IGW to modify the spin frequency of the core is therefore dependent on the rate at which convection generates high frequency waves ($\omega > \omega_c$), i.e., it is dependent on the wave spectrum.

\subsection{Application to the Sun}

In the Sun, low $l$ IGW with frequencies larger than the convective turnover frequency $\omega \gtrsim 5 \omega_c$ can traverse the entire radiative zone, and can reduce differential rotation on short time scales ($T_* \sim 10^7-10^8$ years). This result is in accordance with a series of studies (e.g., Talon \& Charbonnel 2005, Charbonnel \& Talon 2005) examining IGW AM transport in solar-like stars, whose more detailed calculations/simulations found that waves reduce differential rotation within the Sun on similar time scales. This result is not strongly dependent on the IGW wave spectrum generated by convection, nor does it require the existence of a shear-layer oscillation. The solar IGW AM transport time scale of $T_* \sim 10^7-10^8$ years also appears to be consistent with observations of the spin-down of young cluster stars (Stauffer \& Hartmann 1987; Keppens et al. 1995; Bouvier et al. 1997; Krishnamurthi et al. 1997; Barnes 2003)

However, IGW by themselves are unlikely to produce the observed rigid rotation of the solar interior. In the Sun, the IGW that reach $r=0$ have angular frequencies $\omega_* \approx 4 \mu {\rm Hz}$, whereas the rotation rate of the radiative zone is $\Omega_{\odot} \simeq 2.6  \mu {\rm Hz}$. In the absence of other AM transport mechanisms, we may therefore expect to observe differential rotation of order $\Delta \Omega \sim \omega_*/m \sim 4 \mu {\rm Hz}$, in contrast to the nearly rigid rotation which is observed ($\Delta \Omega \ll \Omega_\odot$). We conclude that IGW are capable of performing the bulk of the AM transport required to keep the radiative interior of the Sun synchronous with the convective envelope (in agreement with Talon \& Charbonnel 2005, Charbonnel \& Talon 2005), but that another source of torque is required to enforce rigid rotation (in agreement with Gough \& Mcintyre 1998, Denissenkov et al. 2008).

\subsection{Evolution up the Red Giant Branch}

As stars evolve across the sub-giant branch and up the RGB, their cores become opaque to incoming IGW waves and decouple from the surface convection zone. After decoupling, the cores are able to spin-up as they ascend the sub-giant branch, as observed by Deheuvels et al. 2014. However, asteroseismic measurements of the core rotation rate of stars ascending the RGB (Mosser et al. 2012) indicate that the cores of these stars slowly spin {\it down} as they evolve. IGW on their own are likely incapable of producing this spin down. 

However, IGW are capable of changing the stellar spin down to radii of $r \sim 10^{-1} R_\odot$ (in comparison, the base of the convective zone resides at radii $r \sim 0.75 R_\odot$, while the hydrogen burning shell is located at $r \sim 3\times 10^{-2} R_\odot$). This implies that convectively excited IGW can remove most of the AM from the contracting radiative zone and are able to couple the slowly rotating convective zone with the bulk of the moment of inertia of the radiative zone. We predict that only the inner core (i.e., the inner $\sim 0.2 M_\odot$ comprising the helium core, hydrogen burning shell, and a small fraction of the radiative outer core) of RGB stars rotate rapidly, whereas layers exterior to this can be spun down by IGW. Moreover, our results imply that other AM transport mechanisms (e.g., magnetic torques) need only remove the relatively small amount of AM contained in the inner $\sim 0.2 M_\odot$ of the core in order to allow it to spin down on the RGB. 

Additionally, while on the RGB, the material accreting onto the helium core may have been previously spun down by IGW, meaning the core will not rapidly spin-up as it accretes. This possibility is somewhat dependent on the wave spectrum and so we do not investigate it in detail. However, the impeded spin-up of the core would once again allow other AM transport mechanisms to spin down the inner core with only a relatively small amount of AM transport. Lastly, the IGW could enforce large angular velocity gradients between the inner core (which is unaffected by IGW) and the outer core (which is spun down by the IGW). The IGW-induced shear could then enhance the potency of other AM transport mechanisms. Successful descriptions of AM transport may therefore require the simultaneous interplay between IGW and other sources of torque.

\subsection{Constraints on Wave Excitation and Propagation}

Our results, combined with asteroseismic measurements, may place some constraints on the viability of some surprising results from simulations of IGW wave generation/propagation (see e.g., Rogers et al. 2008, Rogers et al. 2013, Alvan et al. 2014). These authors suggest that a radial damping length increased by $\sim (N/\omega) \gg 1$ better describes IGW attenuation, speculating that non-linear wave-wave interactions may be at play. However, if we use this modified damping length in equation \ref{tau}, we find that low frequency IGW can penetrate all the way into the cores of sub-giants/red giants. These IGW could spin down the cores on short time scales, in contrast with their observed rapid rotation. Thus, this weaker IGW damping appears to be inconsistent with observations. 

We may also be able to constrain the frequency spectrum of the convectively excited IGW. If the wave spectrum is somewhat flat ($a \lesssim 3.5$) as suggested by Rogers et al. 2013, then the AM flux carried by high frequency waves ($\omega \gg \omega_c$) is much greater. This would allow high frequency IGW to change the spin of the cores of sub-giants/red giants on short time scales. In the absence of additional AM transport mechanisms, the IGW would generate differential rotation of order $\Delta \Omega \sim \omega_*$, causing the cores to spin {\it faster} than observed, at $P \sim 2 \pi/\omega_* \sim 2$ days. This scenario seems unlikely, as it would require the presence of an additional AM transport mechanism which would mostly erase the IGW-induced shear, yet allow the smaller degree of observed differential rotation to persist. We find it more plausible that the wave spectrum is steep enough ($a \gtrsim 3.5$) to prevent IGW from significantly altering the spin profile of the g-mode cavity in red giants.

\subsection{Relation to Tidal Theories}

Recent studies (Winn et al. 2010, Dawson 2014) of the tidal evolution of hot Jupiters around main sequence stars have suggested that some observed features of the hot Jupiter distribution can be explained by weak AM transport within the stellar interiors. In particular, these studies have suggested that tides only couple a small piece of the stellar moment of inertia (e.g., a solar-like convection zone) to the planetary orbit. Our results suggest such a decoupling to be extremely unlikely, as IGW can reduce differential rotation on timescales shorter than the ages of the hot Jupiter systems.

\subsection{Clump Stars and High Mass Stars}

Asteroseismic analyses of clump stars (Mosser et al. 2012, see also Tayar \& Pinsonneault 2013) burning helium in their core reveal slower core rotation rates, with rotation periods of $P \sim 100$ days. In clump stars, IGW are excited at the top of the helium burning core and the base of the convective envelope, and both sources of IGW must be included in AM transport calculations. Preliminary results reveal that IGW may be sufficient to couple the core and envelope of clump stars. However, these results are somewhat dependent on the wave spectrum, so we defer a more detailed investigation to future publications.

In massive stars nearing core collapse, the stellar structure becomes complex, with onion-like shells of convective/radiative zones. Although stellar evolutionary timescales become extremely short, the vigorous convection generated by nuclear burning in the cores of these stars generates large fluxes of IGW (Quataert \& Shiode 2012, Shiode \& Quataert 2014). It is therefore likely that IGW play an important role in AM transport for these stars, and we hope to explore this issue in a future paper.

\subsection{Uncertainties}

The main uncertainty involved in IGW AM transport is the spectrum of convectively driven waves. Theoretical predictions suffer from our poor understanding of stellar convection, in particular, the inadequacy of mixing length theory. Moreover, they can be sensitive to frequently discarded factors of order unity (e.g., a change in the value of $\omega_c$ by a factor of $\pi$ in equation \ref{Jstar} will alter wave time scales by orders of magnitude). In turn, results from simulations are difficult to interpret and a detailed physical understanding/justification of their outcomes is often lacking. We hope that future simulations of convectively driven IGW can either confirm or deny current expectations and lead to a genuine understanding of convectively driven IGW dynamics.

\section*{Acknowledgments} 

We thank Ellen Zweibel, Lars Bildsten, Lucy Alvan, Sacha Brun, and Tammy Rogers for useful discussions. We also acknowledge insightful comments from the anonymous referee that helped to improve the manuscript. JF acknowledges partial support from NSF under grant no. AST-1205732 and through a Lee DuBridge Fellowship at Caltech. This research was supported by the National Science Foundation under Grant No. NSF PHY11-25915 and by NASA under TCAN grant number NNX14AB53G.

\end{document}